# Effect of annealing temperature on morphology, structure and photocatalytic behavior of nanotubed $H_2Ti_2O_4(OH)_2$

Min Zhang, Zhensheng Jin*, Jingwei Zhang, Xinyong Guo, Jianjun Yang, Wei Li,

Xiaodong Wang, Zhijun Zhang*

*Key Lab of Special Functional Materials, Henan University, Kaifeng 475001, China*

zhenshengjin@henu.edu.cn (Z. Jin).

**Abstract**

Nanotubed titanic acid ($H_2Ti_2O_4(OH)_2$) was prepared from nanotubed sodium titanate ($Na_2Ti_2O_4(OH)_2$) by an ion exchange reaction in a pH = 1 HCl solution. The effect of annealing temperature on the morphology, structure and photocatalytic behavior of nanotubed $H_2Ti_2O_4(OH)_2$ was studied by means of TEM, XRD, DTG, DSC, BET and ESR. The results showed that nanotubed $H_2Ti_2O_4(OH)_2$ is thermally unstable. Its dehydration consists of two steps. In the first-step dehydration, single-electron-trapped oxygen vacancies (SETOVs) were generated. Accompanying the second-step dehydration, the transition of crystal form from orthorhombic system to anatase took place, at the same time the nanotubes broke. At $T > 300\,°C$, when the SETOV concentration greatly increased, the interaction between SETOV happened. $(V_O^o)_x$ formed could play the role of recombination center of photogenerated $e^-–h^+$ and make the photocatalytic behavior of $TiO_2$ (anatase, obtained from 500 °C-treated nanotubed $H_2Ti_2O_4(OH)_2$) to become bad.

## 1. Introduction

Some alkaline metal titanates with different compositions and morphologies have been reported in literature, such as: tunnel structured $Na_2Ti_{16}O_{13}$ and zigzag structured $Na_2Ti_3O_7$ [1,2], layered $Na_4Ti_9O_{20}·xH_2O$ [3], layered lithium titanate $(Li_{1.81}H_{0.19})Ti_2O_5·2.2H_2O$ [4], fibrous $K_2Ti_4O_9$ [5], etc.

Titanic acids (such as $H_2Ti_3O_7$, $H_4Ti_9O_{20}·xH_2O$, $H_2Ti_2O_5·2.2H_2O$, $H_2Ti_4O_9$, etc.) could be obtained by treating the corresponding titanates with a 0.1–1.0 M HCl or $HNO_3$ [2–5]. At 700 °C, $H_4Ti_9O_{20}$ thermally decomposed to $TiO_2$ [3]:

$$H_4Ti_9O_{20} \rightarrow 9TiO_2 + 2H_2O$$

Layered $(Li_{1.81}H_{0.19})Ti_2O_5·2.2H_2O$ showed a C-base-centered orthorhombic system with the lattice constants $a_0 = 16.66 \pm 0.02\,Å$, $b_0 = 3.797 \pm 0.002\,Å$ and $c_0 = 3.007 \pm 0.003\,Å$. Upon conversion to the hydrogen form the lattice constants changed to $a_0 = 18.08 \pm 0.03\,Å$, $b_0 = 3.784 \pm 0.003\,Å$ and $c_0 = 2.998 \pm 0.002\,Å$, respectively [4]. The photocatalytic properties of $RuO_2$-dispersed $M_2Ti_6O_{13}$ (M = Na, K, Rb, Cs) for water decomposition were investigated by Inoue and coworkers [6,7]. The water photosplitting activity depended on the kinds of alkaline metal atom and increased in the order of Na > K > Rb > Cs. Chatterjee et al. used density functional theory to rationalize the location and activity of ruthenium oxide incorporated into sodium hexatitanates [8].

A new nanotubed material was obtained by the reaction of polycrystalline $TiO_2$ with concentrated NaOH solution at 110 °C. From the contents of Na, Ti and structural water determined, it is concluded that the nanotubed material is $Na_2Ti_2O_4(OH)_2$ [9], rather than $TiO_2$, $TiO_x$ or $H_2TiO_3$ [10–17]. After treatment with a HCl solution of pH = 1, nanotubed $Na_2Ti_2O_4(OH)_2$ can be converted to nanotubed $H_2Ti_2O_4(OH)_2$ [9]. The physico-chemical properties of both nanotubed $Na_2Ti_2O_4(OH)_2$ and $H_2Ti_2O_4(OH)_2$ are known little yet. Recently, we found that after appropriate treatment, nanotubed $H_2Ti_2O_4(OH)_2$ showed both strong visible light absorption and luminescence [18,19]. In this paper, the effect of annealing temperature on the morphology, structure and photocatalytic behavior of nanotubed $H_2Ti_2O_4(OH)_2$ is reported.

---

* Corresponding authors. Tel.: +86-378-2192337; fax: +86-378-2867282.
*E-mail address:* 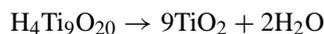 zhenshengjin@henu.edu.cn (Z. Jin).

## 2. Experimental

### 2.1. Preparation of samples [9]

The 300 ml of 40%(w/w) NaOH aqueous solution was placed in a PTFE bottle, equipped with a reflux condenser. Then, the bottle was placed in an oil bath. When the NaOH solution was heated up to 110 °C, 6 g anatase $TiO_2$ powder was added into it and stirred magnetically. After 20 h, the reaction stopped. When the dispersion cooled down to room temperature, it was diluted with de-ionized water to pH = 13.5. Part of the solid settled from the dispersion was washed with anhydrous ethanol for removing the free NaOH adsorbed on both outside and inside surface of the nanotubed material, then filtered and dried under vacuum at room temperature(sample nanotubed $Na_2Ti_2O_4(OH)_2$, Na/Ti atomic ratio = 1). The remained part was washed with de-ionized water to ca. neutral, filtered, then immersed in a pH = 1.0 HCl solution for 5 h and washed with de-ionized water to remove $Cl^-$, and dried under vacuum at room temperature (sample nanotubed $H_2Ti_2O_4(OH)_2$, Na/Ti atomic ratio = 0, $H_2O/TiO_2$ mole ratio close to 1). Heat treatment of samples was carried out in tubular furnace under air flow.

### 2.2. Characterization

Transmission electron microscopic (TEM) patterns were taken on a JEM-2010 electron microscope. X-ray diffraction (XRD) patterns were measured by a X'Pert Pro X-ray diffractometer. Differential thermogravimetric analysis (DTG) and differential scanning calorimetric measurement (DSC) were performed with a SEIKO Exstar-6000 thermal analysis system. BET specific surface areas were determined with a Micromeritics ASAP 2010 apparatus. Diffuse reflectance spectra (DRS) were recorded on a Shimadzu U-3010 spectrometer. Electron spin resonance (ESR) spectra were obtained on a Brüker ESP300E apparatus in air ambience. The g-tensors of ESR signals were obtained by taking g = 2.0036 of diphenyl picryl hydrazyl (DPPH) as reference.

### 2.3. Evaluation of photocatalytic behavior

The photocatalytic behavior of nanotubed $H_2Ti_2O_4(OH)_2$ was evaluated by propylene oxidation removal. The photocatalytic reactor was made of a flat quartz tube, and two 4 W black light lamps ($\lambda$ = 365 nm) were located outside the reactor. The catalyst was spread on both sides of surface-roughened glass plate. Feed gas of 500 ppm(v) propylene was made up of pure $C_3H_6$ and air, and was stored in a high pressure cylinder. The change of $C_3H_6$ concentration before and after reaction was determined by chromatographic method with hydrogen flame detector.

### 2.4. Reagents

Raw $TiO_2$ (anatase, specific surface area = 59 $m^2 \cdot g^{-1}$) was purchased from Hehai Nano-Technology Company, Jiangsu Province; NaOH, analytically pure, product of Tianjin Chemical Reagents Factory; HCl, analytically pure, product of Kaifeng Chemical Reagents Factory; anhydrous ethanol, analytically pure, produced by Tianjin Hongyan Reagents Factory.

## 3. Results and discussion

The morphologies of as-prepared nanotubed $Na_2Ti_2O_4(OH)_2$ and $H_2Ti_2O_4(OH)_2$ are identical, which have a layered structure with the distance between the adjacent layers ca. 0.8 nm (Fig. 1a and b). Both show an orthorhombic

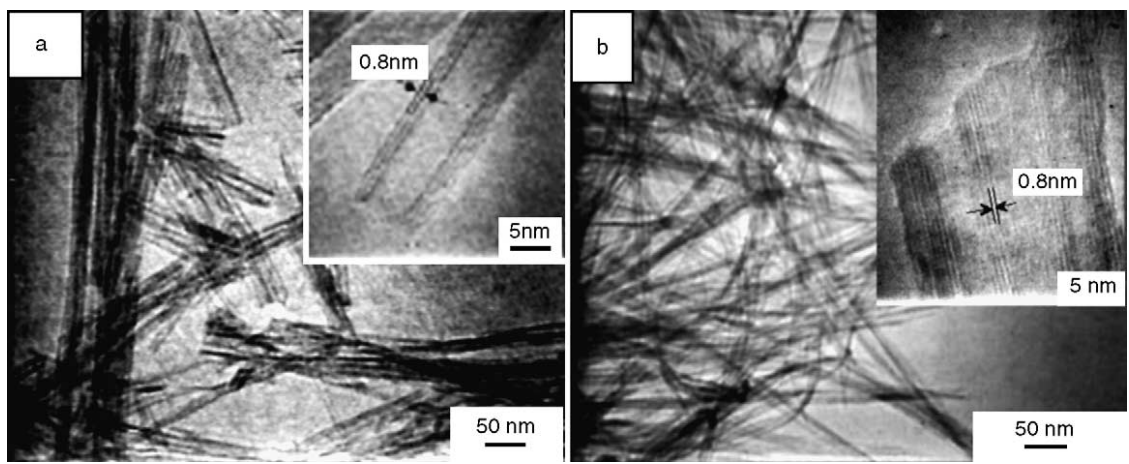

Fig. 1. TEM images of nanotube materials: (a) nanotubed $Na_2Ti_2O_4(OH)_2$ (insert is a four-layered nanotube with inner and outer diameters 6.4 and 9.3 nm, respectively); (b) nanotubed $H_2Ti_2O_4(OH)_2$ (insert is a multi-layered nanotube).

Table 1
XRD data of nanotubed $Na_2Ti_2O_4(OH)_2$

| $2\theta$ | $d$ (Å) | $h\,k\,l$ | $I/I_0$ |
|---|---|---|---|
| 9.18 | 9.63[a] | 2 0 0 | 100 |
| 24.30 | 3.66 | 1 1 0 | 32 |
| 28.14 | 3.17 | 6 0 0 | 60 |
| 34.24 | 2.62 (?) | 3 0 1 | 8 |
| 38.06 | 2.36 (?) | 5 0 1 | 10 |
| 48.14 | 1.89 | 0 2 0 | 69 |
| 61.76 | 1.50 (?) | 0 0 2 | 8 |

Lattice constants: $a_0 = 19.26$ Å, $b_0 = 3.78$ Å.

[a] $d_{200}$ is approximately equal to the distance between adjacent layers (0.8 nm) in Fig. 1a.

system with the lattice constants, $a_0 = 19.26$ Å, $b_0 = 3.78$ Å, but their peak intensities are different, especially for peaks (2 0 0), (1 1 0) and (6 0 0) (Table 1 and Fig. 2) [9,5,20], it may be a result of the replacement of $Na^+$ ions in $Na_2Ti_2O_4(OH)_2$ by $H^+$.

Figs. 3 and 4 indicate the TEM and XRD patterns of nanotubed $Na_2Ti_2O_4(OH)_2$ annealed at different temperatures for 2 h. It is obviously seen that at annealing temperature $(T) \leq 500\,°C$ its nanotube morphology and crystalline form remain unchanged, except the slight shift of $d_{200}$: i.e. 0.963 nm for as-prepared, 0.874 nm for 100–200 °C, 0.740 nm for 300–500 °C (see Fig. 4). The detailed discussion about the formation mechanism and structure of nanotubed $Na_2Ti_2O_4(OH)_2$ has been reported in our former paper [9]. The layered structure shown in Fig. 5a suggests that the decrease of $d_{200}$ is related to the dehydration of interlayered OH groups:

$$Na_2Ti_2O_4(OH)_2 \rightarrow Na_2Ti_2O_5 + H_2O \quad (1)$$

Such a type of dehydration could reduce the interlayer distance but does not destroy the tubular shape. As for nanotubed $H_2Ti_2O_4(OH)_2$, at $T \leq 300\,°C$ there are a little change in nanotube length and (2 0 0) peak intensity (see Figs. 6a–d and 7a–d). But at $T > 300\,°C$, accompanying with the formation of anatase phase [21], nanotubes break (Figs. 6e and f and 7e and f). These results reveal that the dehydration process of nanotubed $H_2Ti_2O_4(OH)_2$ is more complex than that of nanotubed $Na_2Ti_2O_4(OH)_2$. In Fig. 5b ($Na^+$ replaced by $H^+$), there are two types of dehydration: (i) dehydration of intralayered OH groups; (ii) dehydration of interlayered OH groups. At $T \leq 300$ °C the dehydration of intralayered OH groups caused a little change in nanotube length and (2 0 0) peak intensity. When annealing temperature >300 °C, the dehydration of interlayered OH groups induced the change of crystalline form from orthorhombic system to anatase, at the same time, nanotube morphology was destroyed. These two types of dehydration reactions can be formulated as follows:

$$H_2Ti_2O_4(OH)_2 \rightarrow H_2O + H_2Ti_2O_5 \qquad (2)$$

$$H_2Ti_2O_5 \rightarrow H_2O + 2TiO_2(\text{anatase}) \qquad (3)$$

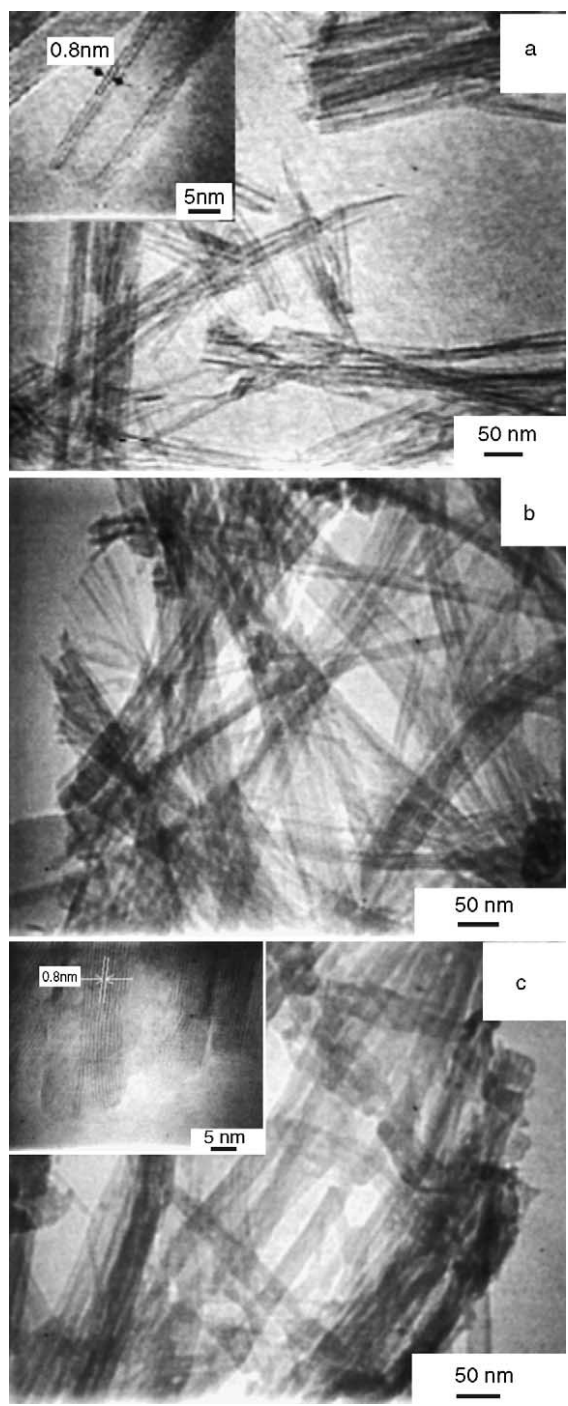

Fig. 3. TEM images of nanotubed $Na_2Ti_2O_4(OH)_2$ annealed in air at different temperatures for 2 h: (a) as-prepared (inset is a multi-layered nanotube); (b) 300 °C; (c) 500 °C (inset is a multi-layered nanotube).

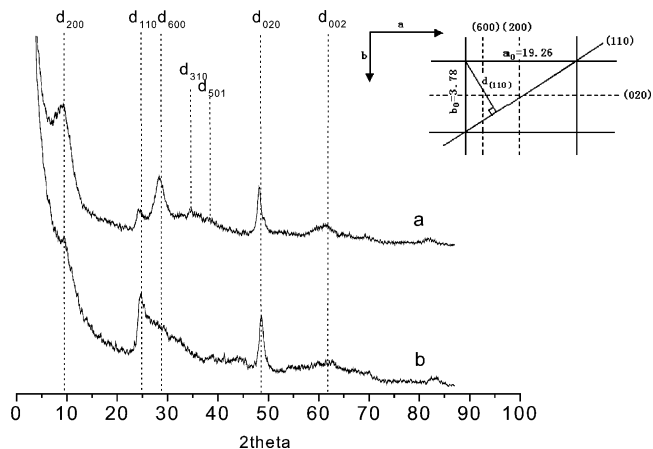

Fig. 2. XRD spectra of nanotube materials: (a) nanotubed $Na_2Ti_2O_4(OH)_2$; (b) nanotubed $H_2Ti_2O_4(OH)_2$; inset: hypothetical schematic view of orthorhombic system, $X$–$Y$ direction.

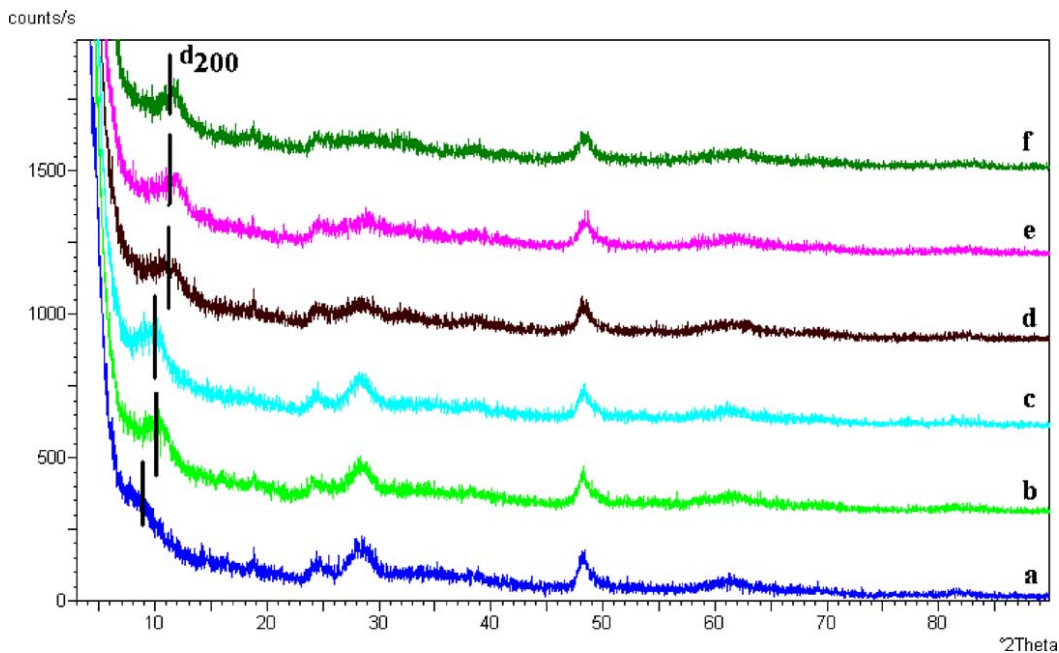

Fig. 4. XRD spectra of nanotubed $Na_2Ti_2O_4(OH)_2$ annealed in air at different temperatures for 2 h: (a) as-prepared; (b) 100 °C; (c) 200 °C; (d) 300 °C; (e) 400 °C; (f) 500 °C.

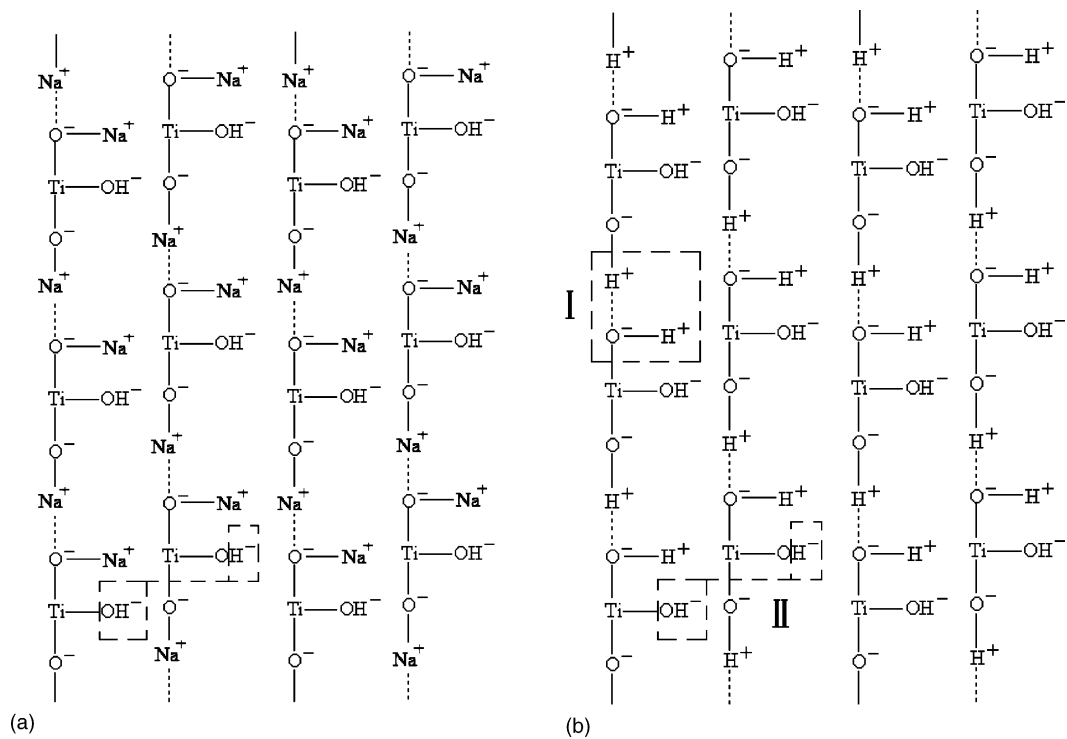

Fig. 5. Schematic diagram of layered composition (X–Y section): (a) nanotubed $Na_2Ti_2O_4(OH)_2$; (b) nanotubed $H_2Ti_2O_4(OH)_2$.

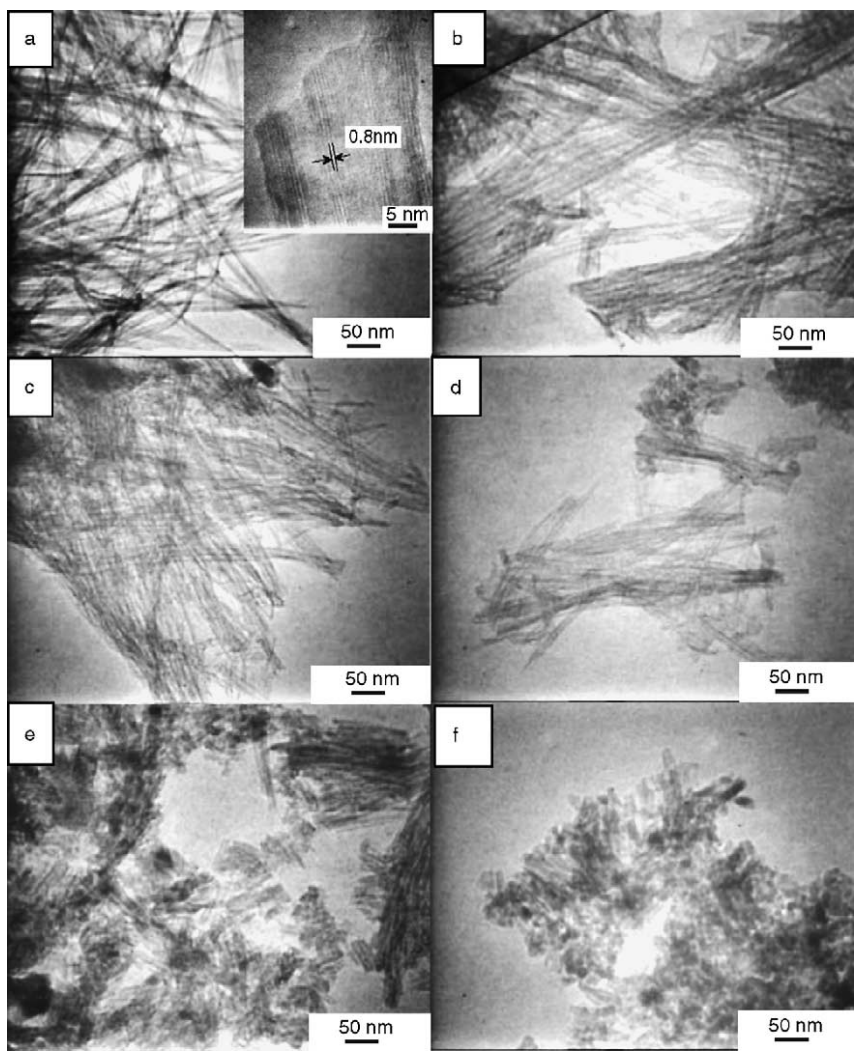

Fig. 6. TEM images of nanotubed $H_2Ti_2O_4(OH)_2$ annealed in air at different temperatures for 2 h: (a) as-prepared (inset is a multi-layered nanotube); (b) 100 °C; (c) 200 °C; (d) 300 °C; (e) 400 °C; (f) 500 °C.

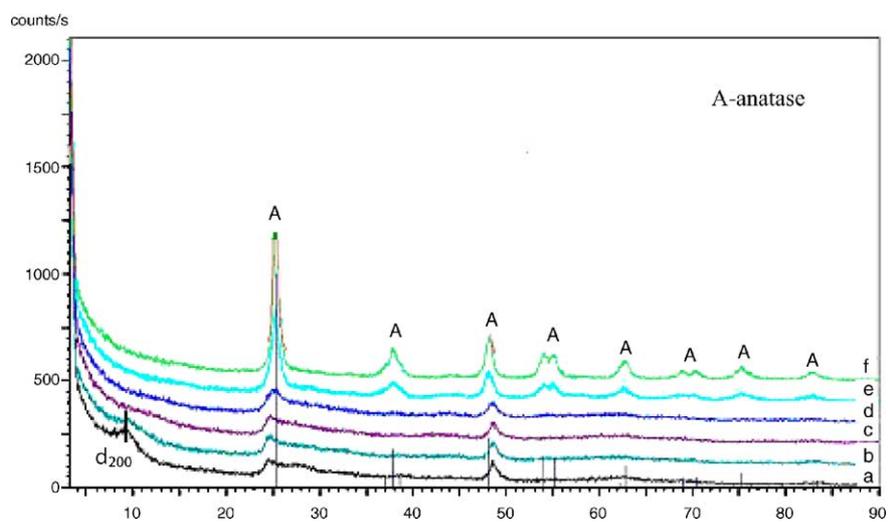

Fig. 7. XRD spectra of nanotubed $H_2Ti_2O_4(OH)_2$ annealed in air at different temperature for 2 h: (a) as-prepared; (b) 100 °C; (c) 200 °C; (d) 300 °C; (e) 400 °C; (f) 500 °C.

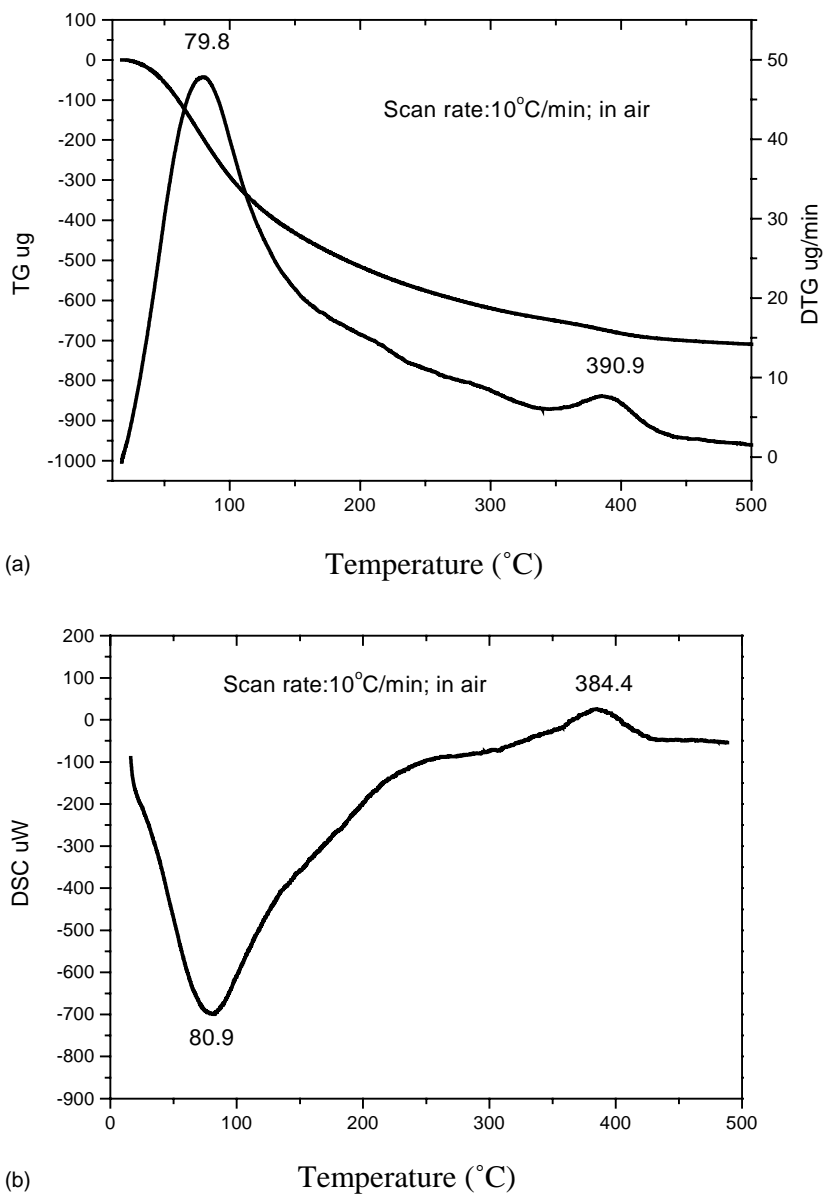

Fig. 8. Thermochemical property of nanotubed $H_2Ti_2O_4(OH)_2$: (a) TG and DTG curves; (b) DSC curve.

DTG, DSC and specific surface area ($S$) data provide further evidences for above two types of dehydration. In Fig. 8a, there are two broad DTG peaks: broad peak I (room temperature to 300 °C, peak at 79.8 °C) consists of the desorption of physically adsorbed water and intralayered water (formula (2)), this peak is corresponding to the endothermic peak on DSC curve (Fig. 8b); broad peak II (300–500 °C, peak at 390.9 °C) represents the desorption of interlayered water (formula (3)), this peak is corresponding to the exothermic peak on DSC curve (Fig. 8b) which reveals that the transition of crystal form from orthorhombic system to anatase is exothermic. Fig. 9 indicates the dependence of specific surface area of nanotubed $H_2Ti_2O_4(OH)_2$ on annealing temperature. In the range of room temperature to 300 °C, S decreases slowly($S_{R.T.} = 402\,m^2\,g^{-1}$, $S_{300} = 354\,m^2\,g^{-1}$); at $T > 300$ °C, the nanotubes break, then $S$ decreases sharply ($S_{400} = 190\,m^2\,g^{-1}$, $S_{500} = 118\,m^2\,g^{-1}$, $S_{600} = 60\,m^2\,g^{-1}$).

DRS spectra (Fig. 10, inset) of both as-prepared nanotubed $H_2Ti_2O_4(OH)_2$ and 500 °C-treated nanotubed $H_2Ti_2O_4(OH)_2$ are similar to raw $TiO_2$, however, their photocatalytic behavior are greatly different. The results shown in Figs. 6–8 have proved that: $T \leq 300$ °C, $H_2Ti_2O_5$ forms (formula (2)), $T > 300$ °C, the transition of crystal form takes place (formula (3)). Fig. 10a shows that the photocatalytic activity of both $H_2Ti_2O_4(OH)_2$ and $H_2Ti_2O_5$ are very low, once the formation of anatase phase begins, the photocatalytic activity increases dramatically and reaches a maximum at ca. $T = 500$ °C (the dependence of photocatalytic activity of nanotubed $Na_2Ti_2O_4(OH)_2$ on annealing temperature was also determined, both $Na_2Ti_2O_4(OH)_2$ and $Na_2Ti_2O_5$ are inert (Fig. 10b) for this reaction). It should be pointed out that the photocatalytic activity of 500 °C-treated nanotubed $H_2Ti_2O_4(OH)_2$ (i.e. $TiO_2$, anatase) is only about 60% of that of raw $TiO_2$ (point c in Fig. 10). For explaining this phenomenon, ESR signals of nanotubed $H_2Ti_2O_4(OH)_2$ treated at different temperatures were measured in air ambience (see Fig. 11). The results show that along with the increase of annealing temperature, the ESR signal varies from a single symmetrical peak ($g = 2.003$) to an unsymmetrical peak ($g_\perp = 2.005$, $g_\parallel = 1.985$), the transition temperature of ESR behavior coincides with that of the crystalline form and morphology shown in Figs. 7 and 6, respectively. From Fig. 5b we can see that the intralayered dehydration of nanotubed $H_2Ti_2O_4(OH)_2$ produce two types of lattice defects: oxygen vacancy and hydrogen vacancy. According to the principle of neutrality and Kröger and Vink's symbols, it may be expressed as follows [26]:

$$2OH \rightarrow H_2O + V_O{}^o + V'_H \qquad (4)$$

where $V_O{}^o$ is the single-electron-trapped oxygen vacancy (SETOV) with one effective positive charge, $V'_H$ is the hydrogen vacancy with one effective negative charge.

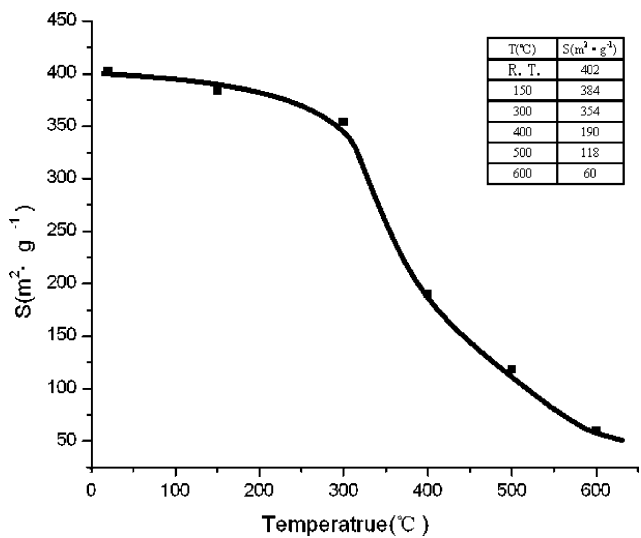

Fig. 9. Specific surface area ($S$) of nanotubed $H_2Ti_2O_4(OH)_2$ annealed at different temperatures for 2 h (as-prepared sample evacuated at room temperature for 20 h).

It was known that oxygen vacancy trapped one electron can be described by a model of a singly-ionized helium-like atom and shows a symmetrical ESR signal with $g = 2.003$ [18,22–25]. Fig. 11b–d evidence the existence of SETOV. At $T > 300\,°C$, when the SETOV concentration greatly increased, the interaction between SETOV happened [26]:

$$xV_O{}^o \rightleftharpoons (V_O{}^o)_x \qquad (5)$$

$(V_O{}^o)_x$ formed shows an unsymmetrical ESR signal with $g_\perp = 2.005$, $g_\parallel = 1.985$ (see Fig. 11e–g), which could play

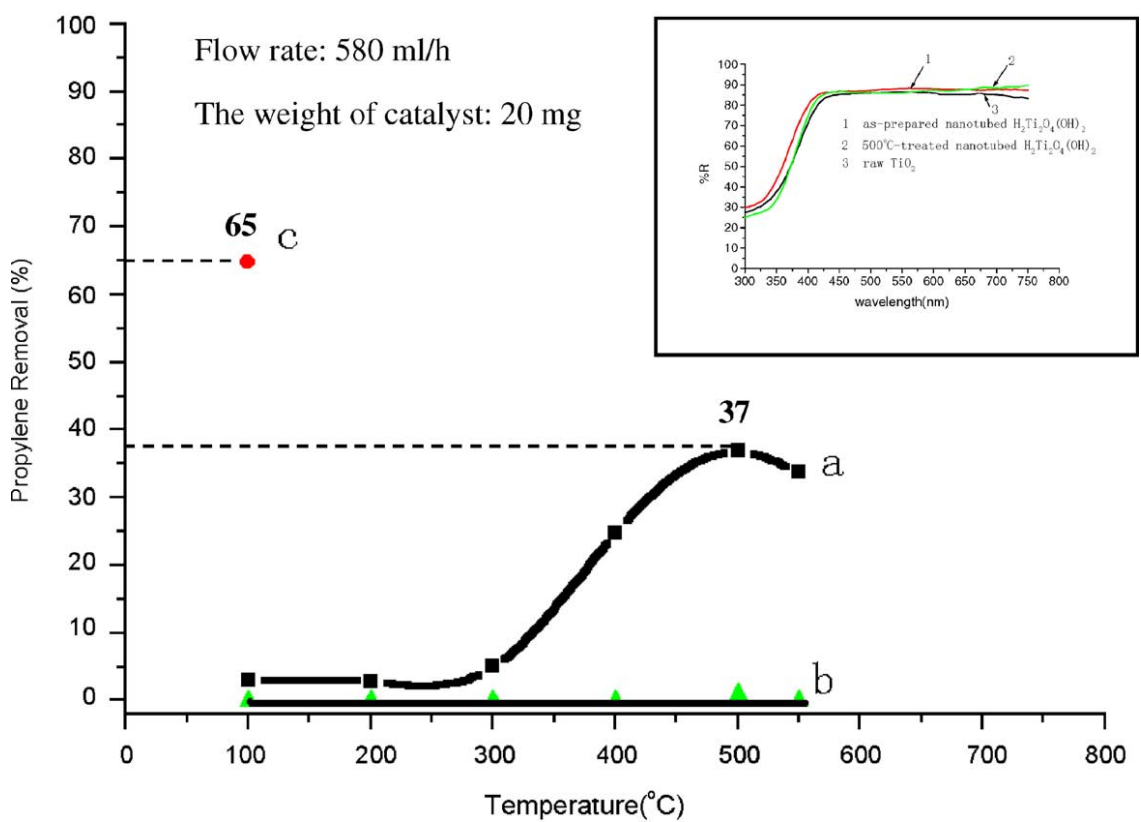

Fig. 10. Dependence of photocatalytic behavior on annealing temperature: (a) nanotubed $H_2Ti_2O_4(OH)_2$; (b) nanotubed $Na_2Ti_2O_4(OH)_2$; (c) raw $TiO_2$ (inset: DRS spectra).

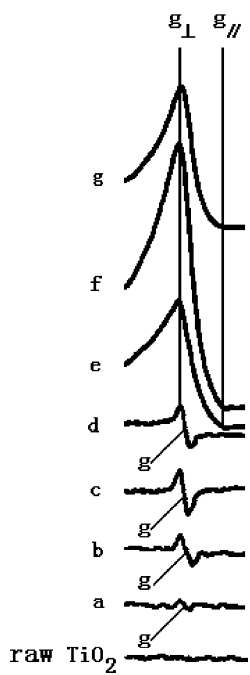

Fig. 11. ESR spectra of nanotubed $H_2Ti_2O_4(OH)_2$ annealed at different temperatures for 2 h: (a) as-prepared; (b) 100 °C; (c) 200 °C; (d) 300 °C; (e) 400 °C; (f) 500 °C; (g) 600 °C ($g = 2.003$, $g_\perp = 2.005$, $g_\parallel = 1.985$).

the role of recombination center of photogenerated $e^-$–$h^+$ and make the photocatalytic behavior of $TiO_2$ (anatase, obtained from 500 °C-treated nanotubed $H_2Ti_2O_4(OH)_2$) to become bad (compared with raw $TiO_2$). As for the decrease of activity at 550 °C (Fig. 10a), it should be resulted from the further decrease of S after the formation of anatase phase (see Fig. 9).

## 4. Conclusions

The morphology and structure of nanotubed titanic acid ($H_2Ti_2O_4(OH)_2$) is thermally unstable. Its dehydration consists of two steps: (i) dehydration of intralayered OH groups at annealing temperature $\leq 300$ °C; (ii) dehydration of interlayered OH groups at annealing temperature $> 300$ °C. In the first-step dehydration, $V_O^o$ are generated. Accompanied with the second-step dehydration, the transition of crystal form from orthorhombic system to anatase takes place, at the same time, nanotubes break and the interaction between SETOV happens. $(V_O^o)_x$ formed could play the role of recombination center of photogenerated $e^-$–$h^+$, and make the photocatalytic behavior of $TiO_2$ (anatase, obtained from 500 °C-treated nanotubed $H_2Ti_2O_4(OH)_2$) to become bad.

## Acknowledgements

We acknowledge the support of the National Natural Science Foundation of China (No. 20071010). The authors are indebted to Porf. Xu Yuanzhi and Chen Jingrong's help for the measurements and discussion of ESR data.